# SHORT GAMMA-RAY BURSTS WITH EXTENDED EMISSION

J. P. Norris and J. T. Bonnell

Astroparticle Physics Laboratory
NASA/Goddard Space Flight Center, Greenbelt, MD 20771.

ABSTRACT

The recent association of several short gamma-ray bursts (GRBs) with early type galaxies with low star formation rate demonstrates that short bursts arise from a different progenitor mechanism than long bursts. However, since the duration distributions of the two classes overlap, membership is not always easily established. The picture is complicated by the occasional presence of softer, extended emission lasting tens of seconds after the initial spike-like emission comprising an otherwise short burst.

Using the large BATSE sample with time-tagged event (TTE) data, we show that the fundamental defining characteristic of the short burst class is that the initial spike exhibits negligible spectral evolution at energies above ~ 25 keV. This behavior is nearly ubiquitous for the 260 bursts with $T_{90} < 2$ s where the BATSE TTE data type completely included the initial spike: Their spectral lags measured between the 25–50 keV and 100–300 energy ranges are consistent with zero in 90–95% of the cases, with most outliers probably representing the tail of the long burst class. We find the same signature – negligible spectral lag – for six Swift/BAT short bursts and one HETE-2 short burst. We also analyze a small sample of "short" BATSE bursts – those with the most fluent, intense extended emission. The same lack of evolution on the pulse timescale obtains for the extended emission in the brighter bursts where significant measurements can be made. One possible inference is that both emission components may arise in the same region. We also show that the dynamic range in the ratio of peak intensities, spike : extended, is ~ $10^4$, and that for some bursts, the extended emission is only a factor of 2–5 less intense. However, for our BATSE sample the total counts fluence of the extended component equals or exceeds that in the spike by a factor of several. A high Lorentz factor, $\Gamma \sim 500–1000$, might explain the negligible lags observed in short bursts.

## 1. INTRODUCTION

In the Konus catalog of short gamma-ray bursts (GRBs), Mazets et al. (2002) discuss "short" bursts which are accompanied by low-intensity, extended emission persisting for tens of seconds after the initial spike-like, harder emission. In the Konus sample the extended component is detected individually in 11 of 130 short bursts. Frederiks et al. (2004) summarize this work and report its presence when tens of bursts are averaged. A similar spectrally soft, extended component was detected in the aggregate for BATSE bursts, and interpreted as the onset of a weak hard X-ray afterglow (Lazzati, Ramirez-Ruiz, & Ghisellini 2001; Connaughton 2002).

Several questions remain about the two components and their causal connection. One of the most fundamental, as mentioned by Mazets et al., is how to distinguish short bursts from the more prevalent long bursts, given that some short bursts have extended emission, whereas some long bursts have only one or two pulse episodes – where is the dividing line? Also, it is unclear if all short bursts have extended emission at some level – what is the dynamic range in intensity of the extended emission relative to the spike emission? Besides being softer, the temporal and spectral character of the extended emission at the pulse level remained to be elucidated – what is the character of the extended emission's pulses, if it can be characterized as such?

The delineation between the two recognized classes of bursts has often been made in a two-dimensional diagram of spectral hardness vs. duration, in which short bursts appear slightly harder than long bursts (Kouveliotou et al. 1993; Meegan et al. 1995). However, the observed average difference in hardness may be at least partially attributable to obvious selection effects: In any GRB trigger algorithm the instrument's effective area as a function of energy limits which bursts may trigger. For instance, Sakamoto (2006) showed that the sample of Konus short bursts has a lower hardness ratio than the BATSE sample. Also, trigger algorithms often use multiple energy bands and accumulation timescales; a sample of long bursts will tend to profit more from the latter measure than can the short burst sample. Recall that the first two seconds of long bursts are on average as spectrally hard as short bursts, after which long bursts tend to soften (Ghirlanda et al. 2004). Moreover, brighter long bursts are harder (Mallozzi et al. 1995). In the BATSE sample we would not have detected some of the long bursts near threshold if they did not commence as spectrally hard. Thus, the distribution of hardnesses in both parent populations is probably much different than in detected samples (cf. X-ray Flashes [XRFs] and GRBs).

In this work, we are primarily concerned with the dichotomy in the duration dimension. In fact, this primary defining characteristic of short bursts is insufficient to cleanly isolate a physical, rather than a one-dimensional, phenomenological class. Long and short bursts' quasi log-normal distributions overlap, defining a valley near two seconds, with some members of each group intruding slightly into the other's duration domain. The degree of intrinsic overlap, and the extent to which extrinsic factors smear the two distributions is not yet established, but the important smearing factors are (at least) four: The difference in signal-to-noise (S/N) levels between the most intense and dimmest BATSE bursts (otherwise identical) gives rise to ~ factor of two range in measured duration (Bonnell & Norris 1997). Distant long bursts ($z \sim 2$–10) detectable by Swift (Lamb & Reichart 2000) are time dilated relative to bursts at $z \approx 1$ by factors of 2-5. For short bursts the relative dilation factors will usually be less than two, assuming those detectable by Swift and HETE continue to lie predominantly at redshifts $z < 1$. Therefore, this effect should increase the separation between the classes. Redshift of the spectral energy distribution operates in the reverse direction of time dilation; the dependence of individual pulse durations and spectral lags on energy is roughly $\sim E^{-0.35}$ (Norris et al. 1996; Fenimore et al. 1995). Thus, the energy band of observation is a factor in measured durations of short bursts, which tend to consist of one to just a few pulses (Norris, Scargle & Bonnell 2001: NSB). Moreover, we do not necessarily detect the shortest bursts, since each instrument algorithm requires a minimum fluence to trigger, introducing truncation effects on the short end of the duration distribution (Lee & Petrosian 1996).

These questions are brought into focus and made more pressing with the recent detections of short bursts by Swift and by HETE-2, two with extended emission. Before the Swift era the distance scale(s) and progenitor type(s) of short gamma-ray bursts (GRBs), those persisting for less than ~ 2 seconds, were still uncertain. Only the isotropic distribution of short GRBs – like that of long GRBs – as revealed by the large BATSE sample, provided a substantive argument in favor of their cosmological distances (Briggs et al. 1996). To date (2005/12/10) ten short bursts have been detected and localized by Swift and HETE-2, and four of these bursts have redshifts determined from their host galaxy or apparent membership in a galaxy cluster. Thus the major question concerning the distance scale for short bursts is qualitatively answered, the redshifts for these bursts being $z = 0.16$, 0.226, 0.258, and 1.8 (Fox et al. 2005; Bloom et al. 2005; Gehrels et al. 2005; Barthelmy et al. 2005a; Gladders & Berger 2005). In two cases (GRBs 050509b and 050724) the burst has been clearly associated with an elliptical galaxy with low star formation rate. The case for coalescing NS-NS or NS-BH binaries as the progenitor population – rather than the massive stellar collapse model for long bursts – is strengthened by an analysis of relatively small error boxes of several short bursts from the pre-Swift era, which finds a high probability of association with old galaxies and old clusters (Gal-Yam et al. 2005). However, some degree of hesitation is introduced by the blue color and spiral nature of the host of GRB 050709 – this may indicate some degree of heterogeneity amongst short burst environments and progenitors (Villasenor et al. 2005; Hjorth et al. 2005; Fox et al. 2005). The importance of making a clear interpretation on class membership is that information on the burst progenitor has been mostly predicated on host galaxy characteristics. Clearer corroboration of class membership from the prompt emission would be helpful.

[We note that not all short bursts may be truly cosmological. The giant flare from SGR 1806-20 could have been detected by Swift to ~ 50 Mpc, implying that eventually ~ 5% of Swift yield for short bursts could be giant SGR events from the local neighborhood (Palmer et al. 2005).]

One intriguing complication to this developing picture is that GRBs 050709 and 050724 have softer, extended low-level emission, lasting for tens of seconds after the initial spike-like pulse. On the other hand, a long burst with one intense pulse at burst onset, separated by many seconds from nearly undetectable low-intensity pulses, might be mistaken for a short burst with extended emission. Figure 1 illustrates two such bursts. In this work we show that these cases can easily be distinguished from most of the "short" bursts, now believed to result from NS-NS or NS-BH mergers. The initial double pulses (T90 durations ~ 1.3 and 1.8 seconds) in both bursts shown in Figure 1 evidence clear spectral evolution (lower energies lagging higher energies), as do all long bursts where counting statistics support a measurement of spectral lag (Norris 2002).

In contradistinction, a large majority of short bursts evidence negligible lag at BATSE energies (NSB). This key discriminant, spectral lag, is examined in § 2 for BATSE short bursts, in particular the set of BATSE bursts with short initial spikes that are followed by extended, low-level emission; for the similar burst, GRB 050709, recently detected by HETE-2; and for six short bursts detected by Swift's BAT. We also examine spectral hardness of the extended component and its dynamic range in intensity and fluence compared to the initial spike. We discuss some ramifications of these findings in § 3 and draw conclusions in § 4.

## 2. BURST CHARACTERISTICS

We noticed early in the BATSE mission (e.g., GRBs 910709 and 921022), while fitting backgrounds to the 4-channel 64-ms DISCSC data, the presence of a small fraction of otherwise short bursts, but with extended lower level emission lasting for tens of seconds after the initial intense, short emission. We will refer to this part of the burst as the "initial spike" – since this is the general appearance when binned on timescales longer than ~ 50-100 ms – but in reality this interval (and generally in short bursts) sometimes comprises a group of a few closely spaced, very short episodes of emission (NSB). At the end of the BATSE mission, we had accumulated about twenty candidates for a putative class that, defined solely in terms of duration, appeared to be an odd mix of short and long burst. The primary criterion for inclusion in this original set was that the initial spike was the dominant feature of the burst. About a quarter of these bursts were easily eliminated from further consideration upon recognition that the later pulses were ordinary, but low intensity, pulses that showed the usual organization in time and energy – evidencing longer decays at lower energies. Another quarter were too dim for profitable analysis. Eight bursts remained, and their temporal and spectral characteristics are examined here. The preparation and analysis procedures employed – including background fitting, duration and spectral lag measurements – are described in previous works (Bonnell & Norris 1997; NSB; Norris 2002). Two remarks on these procedures are particularly pertinent to this work: We have assiduously reexamined our background fits several times for all BATSE bursts, allowing for the possibility of extended emission for all apparently short bursts. Also, where candidates for extended emission were identified, the BATSE signals were closely examined to ascertain that detector-to-detector count ratios were commensurate for initial spike and extended emission, and therefore qualitatively consistent with a common localization.

Time profiles of the remaining eight bursts are illustrated in Figure 2. The common interval is 100 seconds, with the intensity rendered in the logarithm. The interval containing the initial spike is binned at the native 64 ms resolution of the BATSE DISCSC data to allow a uniform comparison of the spike interval for all eight bursts, only four of which have higher time resolution TTE data. The remainder of each time profile is binned to 1.024 s. From top down, the bursts are ordered by decreasing peak intensity of the initial spike as determined on the 64-ms timescale. The ordinate minimum in each panel is 100 counts $s^{-1}$, representing ~ 1+ $\sigma$ fluctuations above background level. Besides the solitary initial spike, the other salient feature is the extended, low-level emission, which tends to peak at levels ~ 30–100 times lower than the initial spike (except for GRB 931222, where the peak-intensity ratio is just a few). In binning up, some interesting spiky detail is lost for the bursts with more intense extended emission. In particular, any evidence for spectral evolution should be examined at the highest possible time resolution. This is addressed in §2.2 for the extended emission in the three most fluent and intense bursts (triggers 1997, 2703, and 5725). We now examine spectral lag, hardness ratio, and duration for the spike emission in our sample of Figure 2, and compare to the distributions of these quantities for short BATSE bursts.

### 2.1 *Spectral Lag Analysis of Spikes*

In NSB a cross correlation function (CCF) approach was used to estimate spectral lags for BATSE short bursts. By fitting the CCF near its peak with a cubic, the native binning can be effectively over resolved by a factor of 2–4. The individual TTE photons were bootstrapped to obtain lag error estimates. We revisited the original analysis to ensure that no errant lag determinations were made. That is, infrequently a secondary, non-central lobe of the CCF can be selected to be fitted in the automated process; secondary lobes arise when more than one significant peak is present. This undesired behavior was eliminated by constraining the process to fit the central lobe, and then confirming visually that this occurred correctly for each of 101 bootstrap realizations for a given burst.

The initial spikes of these eight bursts in Figure 2 are related to the BATSE short burst population in a fundamental way, as are the Swift and HETE-2 short bursts. The vast majority of short bursts have negligible spectral lags. Figure 3, adapted from NSB, depicts peak flux versus lag, the latter measured between the BATSE energy channels 25–50 and 100–300 keV for the 260 bursts for which the TTE data type contained essentially all the burst. The lags for BATSE short bursts, represented with light gray diamonds and associated errors, appear to be distributed mostly symmetrically about zero lag.

In Figure 3 the lags for the initial spikes of the eight bursts of Figure 2 are shown as larger diamonds. The black filled diamonds are for the four bursts that were contained in the TTE data, which we binned to 8 ms to make the lag measurements. For the other four bursts, plotted with open diamonds, we measured the lags using PREB+DISCSC 64-ms data. *These latter four measurements should be regarded as rough estimates, and probably systematically represent upper limits, for the following reason.* While spectral lags in short bursts measured in the BATSE energy bands are usually consistent with zero, these bursts are nevertheless often asymmetric. The longer decays, at significantly lower intensity than the spike emission, develop more noticeably at lower energies. Because of its high intensity, the narrow spike would dominate the lag measurement if measured with sufficiently high time resolution, as in the four bursts measured with TTE data. More simply stated, when binned to 64-ms – longer than pulse durations in many short bursts – the weight of the spike intensity is lost and the low-intensity decay at lower energies asserts a measurable effect. This systematic lag-shifting effect, arising from finite binning on timescales longer than pulse durations, is readily demonstrated in simulations.

The GRB dates for the eight spike-like bursts with extended emission are listed in Table 1, along with the BATSE trigger number, peak flux ($F_p$), duration, lag and associated errors, and the data type used to measure the lag (T = TTE 8-ms, D = 64-ms). The lags should be compared with those of long bursts. Well-determined, significantly positive lags for most bright long bursts span a dynamic range from ~ 300 ms down to ~ 25 ms (see Figure 2b of Norris 2002), contiguous with the lags with the four largest values in Table 1, which also happen to be the ones measured with 64-ms data. Hence, these four bursts could represent the short tail of the long burst distribution, but the systematic effect of the coarsely binned 64-ms data suggests that their lag measurements are effectively upper limits. Thus, as we argue below from additional convergent considerations, these four should probably also be categorized as short bursts.

To see if short bursts in the aggregate really have spectral lags consistent with zero we plot in Figure 4 the distribution of lags in sigmas (1 σ = mean of the plus and minus errors) for the 260 short bursts analyzed in NSB. The two histograms are plotted for the sample divided exactly in half at $F_p = 4.25$ photons $cm^{-2}$ $s^{-1}$. The thick (thin) histogram represents the brighter (dimmer) half. Both subsets are closely distributed about zero lag. However, there is a handful (~ 7) of negative outliers (< –2.5σ), in which upon visual inspection we find that the low energy channel (1) does appear to lead the high energy channel (3). A larger number (~ 20 where 1–2 might be expected) have significant positive lags (> +2.5σ); these may be members of the tail of the long burst distribution. We note that all outliers beyond ±4σ are in the brighter half of the sample (an effort to better characterize these significant outliers is in progress).

In summary, 90–95% of the 260 apparently short bursts analyzed with TTE data have spectral lags consistent with zero; whereas, perhaps ~ 5–7% may actually be "intruders" from the long burst distribution, having perhaps 1–2 close pulses (e.g., Figure 1) which evolve spectrally at BATSE energies, as do all long bursts where a significant measurement can be established (Norris 2002). In fact, the positive outliers mostly inhabit the top fourth (0.3–2.0 s) of the duration distribution for this sample (NSB). Pulses in short bursts tend to be an order of magnitude narrower than pulses in long bursts (e.g., Norris et al. 1994), and this would account for their lags being short, but not negligible: Since short bursts have a pulse width mode of ~ 60 ms (full width at half maximum), the brighter bursts in Figure 3 would evidence significantly positive lags if their $\tau_{lag}/\tau_{width}$ ratios were comparable to those in long bursts.

Thus, negligible spectral lag is the fourth attribute that makes the spike emission in short bursts really different from long bursts, in addition to their narrower pulses, the quasi-separation near 2 seconds in the duration distribution, and their higher hardness ratio in the aggregate. We use the same standard for hardness ratio as Kouveliotou et al. (1993), ratio of total counts in the energy channels 3 and 2 (50–100 keV to 100–300keV), HR3/2. Figure 5 illustrates HR3/2 versus the $T_{90}$ duration for our sample of eight spike bursts with extended emission. The range in HR3/2 is ~ 1–2.6 with an average of 1.63. This is comparable to the average of ~ 1.5 reported in Kouveliotou et al., over a full range of ~ 0.25–3.0. (Their HR3/2 average for long bursts is ~ 0.9). The short burst part of the duration distribution exhibits a broad maximum, ~ 0.2–1.0 s (BATSE 3B catalog; Meegan et al. 1995). The median for the eight durations in Figure 5 is ~ 0.75 s, somewhat higher than for BATSE short bursts generally. However, the Konus sample of 130 short bursts peaks broadly between ~ 0.1–1. seconds (Mazets et al. 2002). Hence, our sample is within the expected dynamic ranges of HR/32 and $T_{90}$ for short bursts.

It might be expected that as the spike duration increases, the lag would increase as well. Figure 6 illustrates duration versus lag for the eight BATSE short bursts with extended emission (open diamonds). There is no trend apparent between these two parameters. Again, the four bursts with apparently positive lags were measured with 64-ms data, and their measurements should be considered as upper limits (see discussion above). For comparison, the lags and initial spike durations for the two long bursts of Figure 1 are also plotted (open circles). As is obvious from Figure 1, their pulses are resolved at 64-ms resolution and the lags are significantly positive. These two bursts represent the few, very shortest long bursts (formally, $T_{90} > 2$ s) with low-level emission after an intense initial spike. However, their "extended" emission appears to have a different character than that of the eight bursts of Figure 2: the later episodes include one well-defined pulse per burst exhibiting the usual spectral evolution.

Also included in Figure 6 are the six short bursts detected by the Swift BAT for which event data is available and the burst was bright enough to make a spectral lag measurement. Table 2 lists the lag and duration measurements for these bursts. In four cases the lag was computed between BAT energy channels 3 and 1 (50–100 keV and 15–25 keV), and in two cases between channels 4 and 2 (100–350 keV and 25–50 keV). Each choice was made to maximize the signal-to-noise level. The channel 4-2 pair corresponds nominally to the BATSE energy channels which were used for the measurements of Table 1. Our studies indicate that comparable lag measurements obtain for channel pairs shifted (quasi-) logarithmically, as is approximately the case here. For the four shortest bursts ($T_{90} \leq 0.25$ s) the lags were measured with the event data binned to 4 ms resolution; for the two longer, dimmer bursts ($T_{90}$ durations 0.6 and 1.2 s) the data were binned to 32 ms. In Figure 6 the results for BAT bursts are plotted with open triangles. All are consistent with zero lag, irrespective of duration.

[The position of one of the six Swift bursts (GRB 050925) lies 0.1° from the Galactic Plane, strongly suggesting that it is a Soft Gamma Repeater (Holland et al. 2005). Ordinary SGR bursts also do not exhibit spectral evolution, although probably for a different reason than the cosmologically distant short burst class (Golenetskii, Il'inskii, & Mazets 1984; Kouveliotou et al. 1987; Norris et al. 1991).]

The $T_{90}$ durations listed in Table 2 were measured using all the BAT event data in the energy band 15–350 keV, rather than burst signal counts generated by the mask-tagged background subtraction procedure. Our approach has the benefit of higher signal to noise, but the background level is less accurately determined. The penalty is small given that there is usually generous latitude for choosing the pre-burst background interval, and in each case we chose the post-burst interval to end before the spacecraft slew commenced. The references in Table 2 give other estimates for the burst durations.

GRB 050709, detected by HETE-2, is similar to our sample of eight BATSE bursts – a short spike with extended emission (Villasenor et al. 2005). We digitized the time profiles of the spike emission in two of HETE-2's energy bands, 30–85 and 85–400 keV, from their Figure 3, and performed the same CCF lag analysis at the 4 ms resolution of these profiles. They report a duration for the spike of 70 ± 10 ms (30–400 keV). The peak flux of the spike in the 50–300 keV (same band as for BATSE bursts) was ~ 20 photons $cm^{-2}$ $s^{-1}$, bright enough to make a good lag measurement (cf. Table 1). The HETE-2 point (lag = $0.0^{+2.0}_{-2.5}$ ms, 1σ formal error) is included in Figure 6 (open square), but with liberal error bars (±3σ) since we are unfamiliar with the particulars of their instrument, and since the HETE-2 energy bands are contiguous and different than BATSE channels 1 and 3 (25–50 and 100–300 keV) which we use for the CCF analysis.

Thus the short bursts from the BATSE and BAT samples and the HETE-2 burst – with or without extended emission – all fit the paradigm for short bursts in general, in that their spectral lags are consistent with zero.

The apparently anomalous part of the picture is the extended emission present in one Swift burst and in the HETE-2 burst, but discernible only in a handful of the ~ 550 short bursts detected by BATSE. We now examine the extended emission for the eight bursts in Figure 2.

### 2.2 *Extended Emission in Short Bursts*

All eight of the BATSE bursts in our sample have $T_{90}$ durations in the range 30–90 s, greater than the median for long bursts. The high counts ratio between the two components – extended to initial spike – makes for the long durations in our BATSE bursts. Because the extended emission component was readily apparent, they were previously assigned membership in the long burst class. However, since the initial pulses have $T_{90}$ < 2 s (Figure 6) in each case followed by a hiatus of several seconds before the extended component commences, the overall appearance is unlike that of most other long bursts. We consider the defining characteristic of the short burst class to be negligible spectral evolution above ~ 20 keV. This property of the initial spikes in our sample leads us to classify them as short bursts. Below we show that the same property obtains for the extended emission in the brightest cases.

Our BATSE sample represents the short bursts with the brightest extended emission, hence those with the highest signal-to-noise level available for study. BATSE's unique combination of longevity (1991–2000) and large area allowed for the detection and examination of these few bursts. We emphasize that the norm for short bursts is very low-level (usually indiscernible) extended emission in an individual burst. Just as the bright end of the log [$N(F_p)$] – log [$F_p$] distribution for GRBs becomes well populated only with sufficient exposure, our sample with intense extended components is available due to the nine years of the Compton GRO mission. In fact, all short bursts might have extended emission, but it is usually indistinguishable from background in individual bursts. Lazzati et al. (2001), Connaughton (2002), and Frederiks et al. (2004) did detect this component in the aggregate time profiles, and reported that this persisted for tens of seconds and was softer than the initial spike.

The Swift and HETE-2 short bursts with extended low-level emission are similar in that the extended emission is softer than the initial spike (Barthelmy et al. 2005a; Villasenor et al. 2005). Mazets et al. (2002) also report a soft extended component (see their Table 3). The same is true for all eight bursts in our sample. Table 3 lists the ratio of total counts > 25 keV of the two components (spike : extended), the HR3/2 measurements for spike and extended emission, and the ratio of the two HR3/2's. The extended emission is always spectrally softer than the spike.

The dynamic range of intensity of the two components is an important quantity for theories to entertain. In our eight BATSE bursts, the extended emission is relatively strong, with peak intensities only a factor of 2–10 lower than the spike emission in three cases (compare Figures 2 and 7). The two brightest extended components attain count rates of $\sim 10^4$ $s^{-1}$. Assuming a typical background rate of 8000 $s^{-1}$, a 1-$\sigma$ fluctuation in Figure 2 is ~ 100 $s^{-1}$. Lazzati et al. (2001) detect the extended component in a sample of 76 co-added bursts with the averaged time profile binned to 16-s samples. In their Figure 1 a 1-$\sigma$ fluctuation translates to $\sim (8000\ s^{-1} \times 16\ s \times 76)^{1/2} \approx 3200$, or 2.5 $s^{-1}$ $burst^{-1}$ in the average. Thus the dynamic range in intensity for the extended component amongst BATSE bursts is > 4000, since some of the bursts in the Lazzati et al. sample must be dimmer than the average. At least 30 BATSE short bursts with $F_p > 10$ photons $cm^{-2}$ $s^{-1}$ show no individual evidence of extended emission, whereas three of our sample have comparable peak fluxes (see Figure 3). The initial spike in GRB 931222 is ~ 2 times more intense than the peak in the extended emission. We conclude that the dynamic range in the ratio of peak intensities, spike : extended is $\sim 10^4$.

The range in the ratio of total counts is also large. For our BATSE sample, the dynamic range of the ratio, spike : extended, is ~ 50:1 (see Table 3). The HETE-2 burst, GRB 050709, also exhibited more than an order of magnitude more counts in the extended emission at low energies, 2–25 keV (Villasenor et al. 2005). However, the initial spike sometimes dominates. The KONUS experiment on the Wind spacecraft detected 11 (out of 130 total) short bursts which individually exhibited extended emission (Mazets et al. 2002; Frederiks et al. 2004). Mazets et al. report fluences for the two components. The fluence ratios tend to be a factor of several lower than our count ratios in Table 3, even after taking into account the difference in hardness ratios of the two components. Since in most bursts the extended emission is detected only in the aggregate, the same reasoning as described above for the range in flux ratio applies for the overall dynamic range of the total counts (or fluence) ratio, which must also be of order $10^4$. Thus physical models must account for how the division in the energy budget between the two components can vary so greatly from burst to burst.

The extended component is visible in only ~ 2% of BATSE short bursts, but it appears that Konus, Swift, and HETE-2 may be detecting a significantly larger fraction of this softer emission. At least two factors may contribute to the higher yields of these two missions. The BATSE Large Area Detectors were sensitive down to ~ 25 keV, and viewed the background over a large portion of the sky. The response for Konus continues to ~ 12 keV, Swift's BAT to ~ 15 keV (Barthelmy et al. 2005b), and to 2 keV for HETE-2 (Ricker et al. 2003). For Swift, the mask-tagged background subtraction method applied to event data yields a time profile with effectively zero signal (but non-zero variance) integrated over directions different than the burst direction, allowing sensitive searches for extended hard X-ray emission from bursts.

An inspection of Figure 2 reveals that the extended emission usually does not commence directly succeeding the spike. Rather, after a hiatus of ~ 10 s the low-level emission builds later to a relatively broad maximum on a timescale of 30–50 s. When the count statistics support examination at 64-ms resolution, the emission is very spiky, as shown in Figure 7 for the three bursts (triggers 1997, 2703, and 5725) with the most intense extended emission. We carried out the CCF analysis for these bursts using the intervals bounded by pairs of dashed lines in Figure 2 (the other three bursts were too weak to give constraining results). Outside the selected intervals, the emission drops off to successively lower valleys between spiky episodes; inclusion of these lower intensity regions tends to increase the error estimates. The results are summarized in Table 4. The spectral lags for extended emission in two bursts are $16^{+12}_{-8}$ ms and $12^{+16}_{-6}$ ms, formally 2σ different from zero lag, and consistent with zero for one burst. The central values for the positive lags are comparable to or less than the lags measured for spikes in four bursts of our sample (compare with Figure 3 and Table 3). Determining whether or not the extended emission in such bursts would usually be consistent with zero when resolved at finer time resolution – as we found to be often the case for the spike emission – will require, e.g., bright Swift bursts examined using time and energy tagged event data. In the case of GRB 050709, Villasenor et al. (2005) report no spectral evolution for the extended emission.

Finally, we ask if there is a relationship between strength of the extended emission and its spectral hardness, relative to the spike emission. Figure 8 illustrates the total counts ratio versus the ratio of the HR3/2's. The hint of a possible trend is apparent: when the extended emission is relatively less fluent, it may tend to be spectrally softer. A larger sample would be necessary to support this loose correlation.

## 3. DISCUSSION

In summary, our spectral lag analysis for eight BATSE short bursts with extended emission shows that a fundamental defining characteristic of the short burst class is that the initial spike exhibits negligible spectral lag at energies above ~ 25 keV. This behavior is nearly ubiquitous for the 260 bursts with $T_{90} < 2$ s where the BATSE TTE data type completely included the initial spike: Their spectral lags measured between the 25–50 keV and 100–300 energy ranges are consistent with zero in 90–95% of the cases. Most outliers have positive lags and probably represent the tail of the long burst distribution. Similarly, our analyses of six Swift BAT short bursts and one HETE-2 burst (Villasenor et al. 2005) show that their initial spikes exhibit negligible spectral lag. Extended emission was evident in the HETE-2 burst (GRB 050709; Villasenor et al. 2005), and one of the Swift bursts (GRB 050724; Barthelmy et al. 2005a). Thus membership for short bursts with extended emission – which can sometimes extend their formal $T_{90}$ measurements to > 2 s – appears decided within the short burst class.

The extended emission in our sample of eight BATSE short bursts is always softer than the spike emission, as was reported for all such bursts observed by HETE-2, Swift, and Konus (Villasenor et al. 2005; Barthelmy et al. 2005a; Mazets et al. 2002), and for averaged time profiles of BATSE short bursts (Lazzati et al. 2001; Connaughton 2002). Considering all these studies combined with our results, we find that the dynamic ranges in both flux and fluence ratios, spike : extended, is very large, of order $10^4$. How the apportioning of the energy budget may vary so greatly between the two components is a major question. We also find that when the extended emission is sufficiently intense to make a measurement, its appearance is very spiky and the average spectral lag is negligible, similar to that of the initial spike.

The resulting canonical picture of “short hard” bursts is modified in several ways. First, when accompanied by sufficiently fluent extended emission, short bursts are not always shorter than ~ 2 s; they can have $T_{90}$ durations lasting tens of seconds. Second, as detailed in the introduction, the appearance that short bursts are spectrally harder than long bursts may be partially or completely attributable to spectral and temporal trigger selection effects. Third, when sufficiently bright to support examination on short timescales, the extended emission has a spiky pulse-like appearance, and thus may not be a standard “afterglow.” Instead, these pulses might be analogous to the X-ray flares observed accompanying the prompt portions of afterglows as revealed by Swift’s XRT, and attributed to central engine activity (Zhang et al. 2005). Fourth, in many cases where the extended component is discernible in an individual burst (Figures 2 and 7; Villasenor et al. 2005; Barthelmy et al. 2005) there is a hiatus – an interval of negligible or low-intensity emission – after the initial spike and before the extended emission strengthens appreciably. This frequently occurring feature has no obvious analog in long bursts.

However, the hiatus is not always present, as can be appreciated from the Konus Catalog of Short Bursts (Mazets et al. 2002). The absence of a hiatus in individual bursts and in the average profile of BATSE short bursts (Lazzati et al. 2001) may be evidence against models involving ejecta from the source impinging on a companion star (see below).

It remains to explain the negligible spectral lags that are characteristic of short bursts at energies above ~ 15 keV, in contrast to the signature of positive lags in long bursts (Norris 2002). Recall that pulses in short bursts are ~ 10–20 × shorter than in most long bursts (NSB). A lower limit on pulse duration due to relativistic beaming is $\tau_{rel} \sim (1+z)R/2c\Gamma^2$ (angular spreading, or the "curvature effect"), where R is the distance of the emission region from the source, and $\Gamma$ is the bulk Lorentz factor of the emitting matter (Fenimore et al. 1996; Sari & Piran 1997). Later emission that arrives during the interval $\tau_{rel}$ from further off axis is lower frequency – this is one inescapable component of spectral lag. In short bursts $\Gamma$ must be sufficiently large so that $\tau_{rel}$ is negligible compared to our instruments' measurement capabilities. In fact, $\tau_{rel}$ must be significantly shorter than the pulse timescale, or else we would be able to measure its contribution to spectral lag in the brighter short bursts (as discussed in § 2.1). Therefore the (dominant) timescale of pulses in short bursts must be attributable to variability of central engine activity, rather than to relativistic beaming. Then to avoid a significant contribution to spectral lag from $\tau_{rel}$, the Lorentz factors in short bursts must be several times higher than in long bursts.

Assuming that $\tau_{rel}$ is a major contribution to lag in both long and short bursts, we can estimate roughly the average $\Gamma$ in short bursts relative to that in long bursts, as follows. Above a peak flux threshold of $F_p > 10$ photons cm$^{-2}$ s$^{-1}$ the lag measurements for both samples are usually accurately measured. The median lag for the 90 BATSE long bursts above this threshold is ~ 48 ms (Norris 2002); and for the same threshold the average lag for 30 BATSE short bursts (excluding the few outliers deemed to be long burst intruders) is 0.1 ± 3.0 ms (1 $\sigma$). Since their lag distribution is near Gaussian, we take the 2-$\sigma$ sample error, $2\times3.0 \text{ ms} / (30-1)^{1/2} \approx 1$ ms, to be the effective upper limit for the short burst sample lag. If the other factors contributing to $\tau_{rel}$ (R and 1+z) were comparable for long and short bursts, then we would have

$$\Gamma_{short} / \Gamma_{long} \sim (\tau_{rel_long} / \tau_{rel_short})^{1/2} > 6.8 . \qquad (1)$$

To take into account the different (observed) redshift distributions of the two classes, the rough estimate in eq. (1) should be reduced by $\delta \sim [(1+z_{long}) / (1+z_{short})]^{1/2}$. Suppose the median short and long burst redshifts are $z_{short} \sim 0.5$ and $z_{long} \sim 2.5$. Then with $\delta \sim 1.5$, $\Gamma_{short} / \Gamma_{long} > 4.5$. Some estimates for $\Gamma_{long}$ are of order 100–200 (e.g., Zhang et al. 2005). Then, modulo the factor due to the emission distances from the source, $(R_{long}/R_{short})^{1/2}$, we would have $\Gamma_{short} \sim 500$–1000.

Similar Lorentz factors are predicted from modeling with general relativistic hydrodynamic codes for outflows from compact object mergers (Aloy, Janka, & Muller 2005). Note that for R = $3 \times 10^{13}$ cm, $\Gamma = 10^3$, and z = 1, angular spreading gives $\tau_{rel}$ = 1 ms.

At energies sufficiently lower than the BATSE bandpass (< 25 keV), short burst pulses may exhibit longer decays – and therefore significant lags – due to the curvature effect. A good example is GRB 050709, where the initial spike emission clearly becomes asymmetric below 10 keV, as revealed by HETE-2 (see Figure 3, Villasenor et al. 2005).

Like the spike emission, the bright, fluent extended emission in three bursts exhibits negligible spectral lag. The general inference is that both emission components may arise in the same region, but (possibly) from different mechanisms: If the extended component were generated at a significantly larger distance from the source than the spike component, then energy-dependent arrival time dispersion would probably be manifest, due to the kinematic correlation between energy and off-axis angle in relativistic beaming.

MacFadyen, Ramirez-Ruiz, and Zhang (2005) propose a model for X-ray flares in burst afterglows, relying on the interaction of the relativistic outflow from an accreting neutron star and a non-compact stellar companion at a separation of approximately a light-minute. Such a scenario clearly has some properties consistent with the extended emission examined here, peaking ~ 30–50 s after the initial spike. For such a binary progenitor, a hiatus between spike and extended emission as well as flux in the extended emission would be expected to correlate with separation between the neutron star and stellar companion. Absence of a hiatus, as apparent for some bursts in the KONUS Catalog of Short Bursts (Mazets et al., 2002), could rule out this model for short bursts in general. Constrained by the binary separation, the interaction region would be compact enough to exhibit negligible geometrically induced lags at gamma-ray energies. The observed spiky structure of the extended emission would suggest a patchy outflow with a filling factor ~ 50% to produce constituent pulses of width ~ 250–1000 ms, as observed.

## 4. CONCLUSIONS

We have presented strong evidence – including results from BATSE, HETE-2 and Swift samples – that as a class, short GRBs have negligible spectral lag at energies above ~ 15–25 keV. The average lag (25–50 keV to 100–300 keV) for the 30 brightest BATSE short bursts is ~ 0.1 ± 0.5 ms, compared to ~ 50 ms for bright long bursts. Lorentz factors several times higher than inferred for long bursts would obtain for short bursts ($\Gamma \sim 500$–$1000$) if a major portion of spectral lag is attributable to relativistic beaming. However, the existence of lags when lower energy bands are considered is not precluded (for example, GRB 050709; Villasenor et al. 2005) since the angular dependence of the "curvature effect" produces longer delays at lower energies.

We show that the extended component – always softer than the initial spike – can manifest a dynamic range in intensity and fluence of order ~ $10^4$ compared to the initial spike. Infrequently, as in our BATSE sample of eight bursts, the strength of the extended emission converts an otherwise short burst into one with a duration that can be tens of seconds, making it appear to be a long burst. How the short burst mechanism can give rise to such a range in total energy in the extended emission is unclear.

We also noted that spectral hardness is not an uniquely defining characteristic for short bursts. Their extended emission is softer, whereas long bursts – at onset as hard as short burst's initial spikes – tend to soften as the burst progresses. Also, a specific instrument's trigger criteria impose strong selection effects, masking the actual distribution of spectra in a parent population. For instance, XRFs were essentially nonexistent in the BATSE sample, but constitute a third of the HETE-2 sample (Sakamoto et al. 2005). Thus, "short hard" bursts are neither necessarily short, nor do they deserve to be labeled as strictly harder than long bursts. We suggest that the current popular nomenclature for the two classes , SHB (for short hard burst) and LSB (for long soft burst), is at best misleading.

However, short bursts do have pulses which are ~ 10–20 × shorter than pulses in long bursts (NSB). The second clear distinction between the two classes that we have emphasized here is spectral lag. These two properties of the prompt emission in short bursts may enable us to discern which class a given burst belongs to, especially when host type, immediate environment, and afterglow information are indecisive.

We are grateful to the anonymous referee for suggestions and corrections that significantly contributed to the substance and organization of this work. We thank Neil Gehrels and Bing Zhang for helpful conversations and encouragement.

TABLE 1. BATSE Bursts: Initial Spike Properties

| GRB Date | Trig# | Fp ($cm^{-2}$ $s^{-1}$) | Dur (s) | $\tau_{lag}$ (ms) | $\varepsilon_{lag}$ (ms) | Data Type |
|---|---|---|---|---|---|---|
| 910709 | 503 | 5.2 | 0.896 | 24.0 | +6.0 -12.0 | D |
| 920525 | 1626 | 3.2 | 0.528 | 26.0 | +12.0 -8.0 | D |
| 921022 | 1997 | 40.3 | 0.384 | 12.0 | +0.0 -2.0 | D |
| 931222 | 2703 | 3.9 | 1.776 | 1.0 | +10.5 -7.5 | T |
| 951007 | 3853 | 3.3 | 1.392 | -6.0 | +35.0 -29.5 | T |
| 961225 | 5725 | 11.7 | 1.216 | 0.5 | +3.0 -2.5 | T |
| 990712 | 7647 | 24.1 | 0.648 | 24.0 | +4.0 -6.0 | D |
| 000107 | 7936 | 1.7 | 0.368 | 6.0 | +20.0 -17.5 | T |

TABLE 2. Swift/BAT Bursts: Initial Spike Properties

| GRB Date | Dur (s) | $\tau_{lag}$ (ms) | $\varepsilon_{lag}$ (ms) | Chan Pair | References |
|---|---|---|---|---|---|
| 050509b | 0.036 ± 0.008 | 4.3 | +3.2 -3.0 | (31) | Gehrels et al. (2005) |
| 050724 | 2.612 ± 0.088 | -4.2 | +8.2 -6.6 | (31) | Barthelmy et al. (2005a) |
| 050813 | 0.564 ± 0.054 | -9.7 | +14.0 -11.0 | (42) | Sato et al. (2005) |
| 050925 | 0.128 ± 0.006 | 8.6 | +6.0 -6.6 | (31) | Markwardt et al. (2005) |
| 051105a | 0.080 ± 0.012 | 6.3 | +5.3 -4.8 | (31) | Cummings et al. (2005) |
| 051210 | 1.368 ± 0.040 | -5.3 | +24.0 -22.0 | (42) | Barthelmy et al. (2005c) |

TABLE 3. Extended Emission Compared to Initial Spike

| GRB Date | Counts Ratio (S/E)* | HR3/2 Spike | $\varepsilon_{HR}$ | HR3/2 Extended | $\varepsilon_{HR}$ | R-HR[+] Ext/Spk | $\varepsilon_{R\text{-}HR}$ |
|---|---|---|---|---|---|---|---|
| 910709 | 0.815 | 1.725 | 0.052 | 0.840 | 0.102 | 0.487 | 0.061 |
| 920525 | 0.500 | 1.025 | 0.061 | 0.748 | 0.097 | 0.730 | 0.104 |
| 921022 | 0.077 | 1.333 | 0.036 | 0.773 | 0.012 | 0.580 | 0.018 |
| 931222 | 0.023 | 1.172 | 0.097 | 1.070 | 0.019 | 0.913 | 0.077 |
| 951007 | 0.461 | 1.475 | 0.084 | 0.421 | 0.079 | 0.285 | 0.056 |
| 961225 | 0.197 | 2.001 | 0.072 | 1.140 | 0.043 | 0.570 | 0.030 |
| 990712 | 0.902 | 2.617 | 0.033 | 0.906 | 0.041 | 0.346 | 0.016 |
| 000107 | 0.397 | 1.677 | 0.236 | 1.298 | 0.277 | 0.774 | 0.198 |

* Ratio of emission in the spike to the extended emission.

+ Ratio of the extended emission hardness ratio to the spike emission hardness ratio.

TABLE 4. Extended Emission Spectral Lag

| GRB Date | $\tau_{lag}$ (ms) | $\varepsilon_{lag}$ (ms) | Interval (s) |
|---|---|---|---|
| 921022 | 16.0 | +12.0 -8.0 | 18.3-37.6 |
| 931222 | 8.0 | +20.0 -22.0 | 18.9-48.8 |
| 961225 | 12.0 | +16.0 -6.0 | 16.4-29.3 |

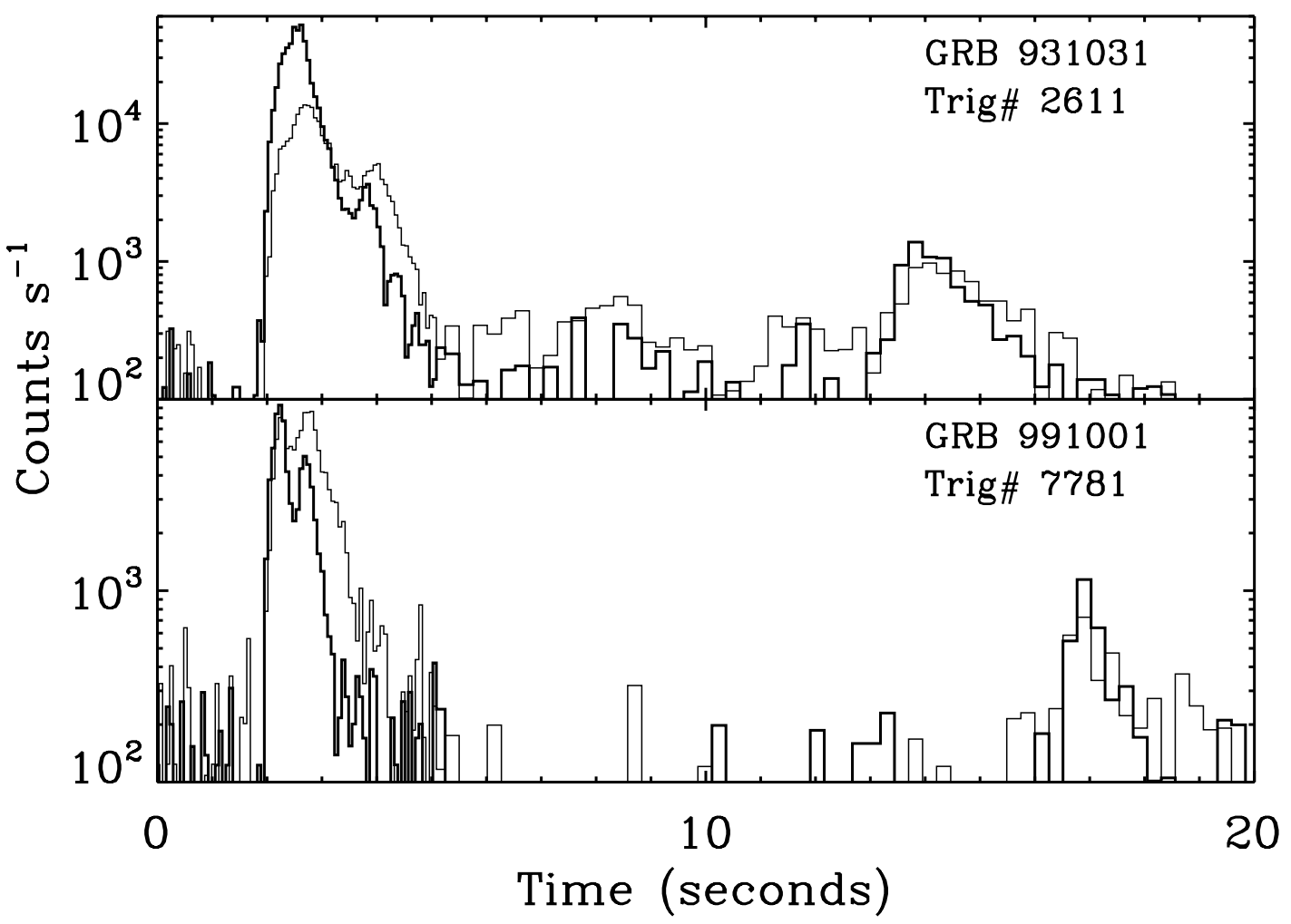

Fig. 1–Time profiles on a logarithmic intensity scale for two BATSE long bursts where the initial, intense emission has a duration $T_{90} < 2$ s, and subsequent low-level emission extends the burst to ~ 20 s. The bin scale is 64 ms for the first 5 s, then 1.024 s. Thick line: 100–300 keV; thin line: 25–50 keV. The usual spectral evolution observed in long bursts is present in both the initial and extended parts of both bursts.

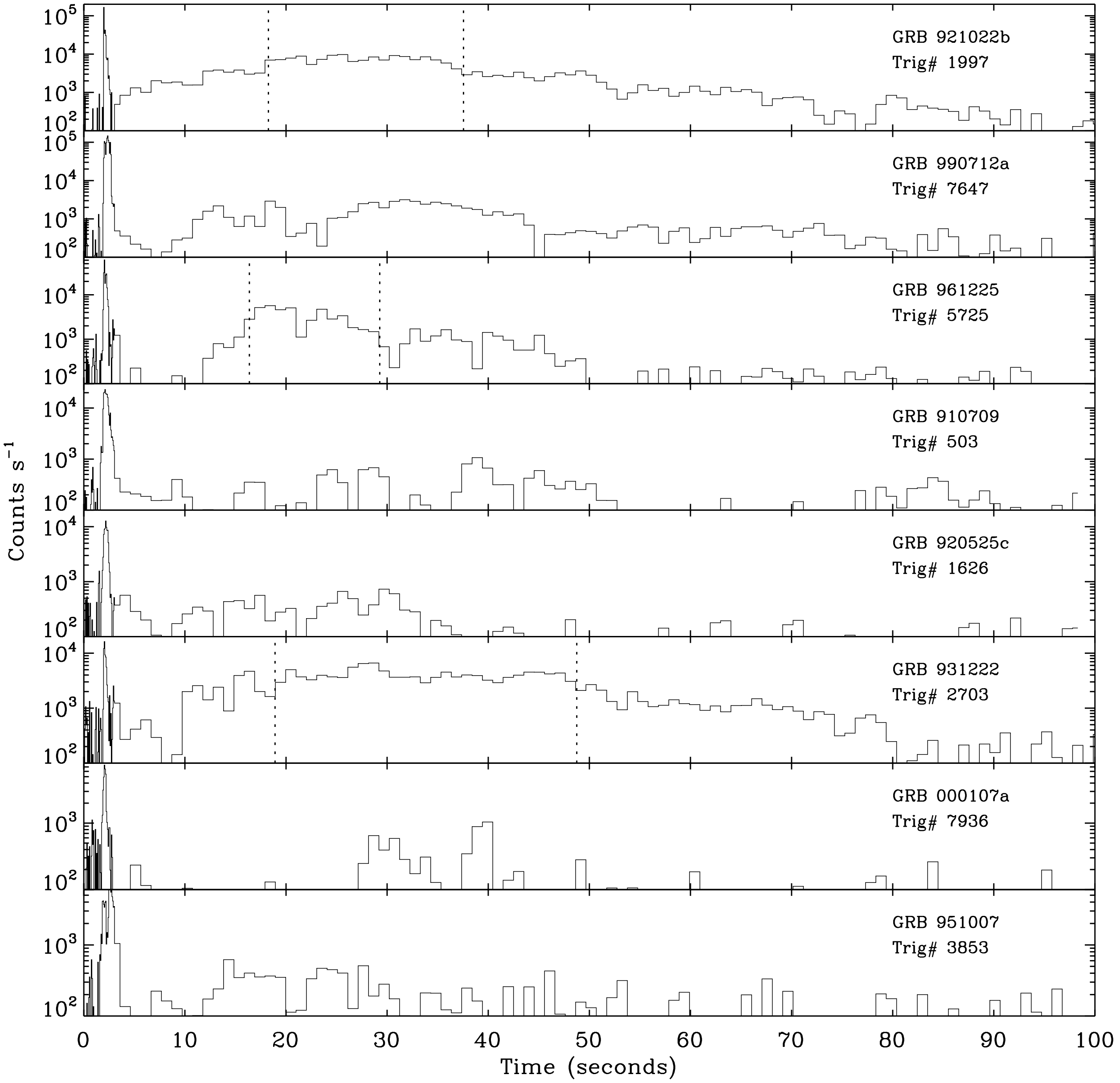

Fig. 2–Similar to Figure 1: time profiles for the eight BATSE spike-like bursts analyzed in this work. Extended emission continues for up to ~ 100 s. The bursts are arranged top to bottom in order of decreasing intensity of the initial spike emission, which exhibits very little or negligible spectral evolution at BATSE energies. The tendency is evident for a hiatus to occur directly after the initial spike, followed by the rising extended emission which is most prominent tens of seconds later. Vertical dotted lines in three profiles indicate intervals of extended emission analyzed for spectral evolution in § 2.2.

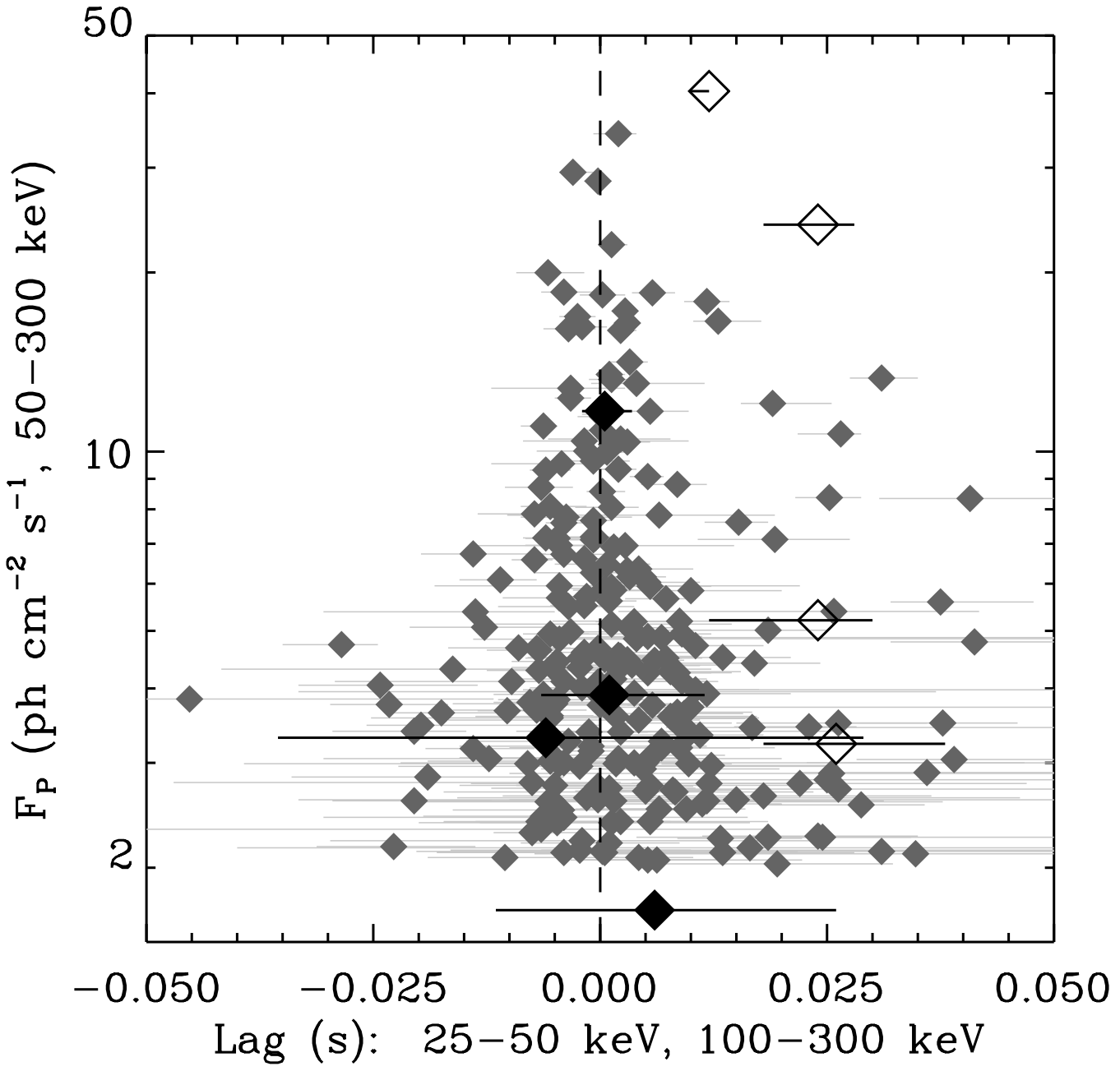

Fig. 3–Peak flux (50–300 keV) vs. spectral lag (25–50 keV to 100–300 keV) for those 260 BATSE short bursts (gray diamonds) with Time Tagged Event (TTE) data which contained the burst. In addition, the lags for the spike emission for the bursts of Figure 2 are plotted with filled black diamonds (TTE data binned at 8 ms) or open diamonds (64-ms PREB+DISCSC data). Values for the latter set should be regarded as upper limits (see text). All (asymmetric) error bars are ±1σ. Adapted from Norris, Scargle, & Bonnell (2001).

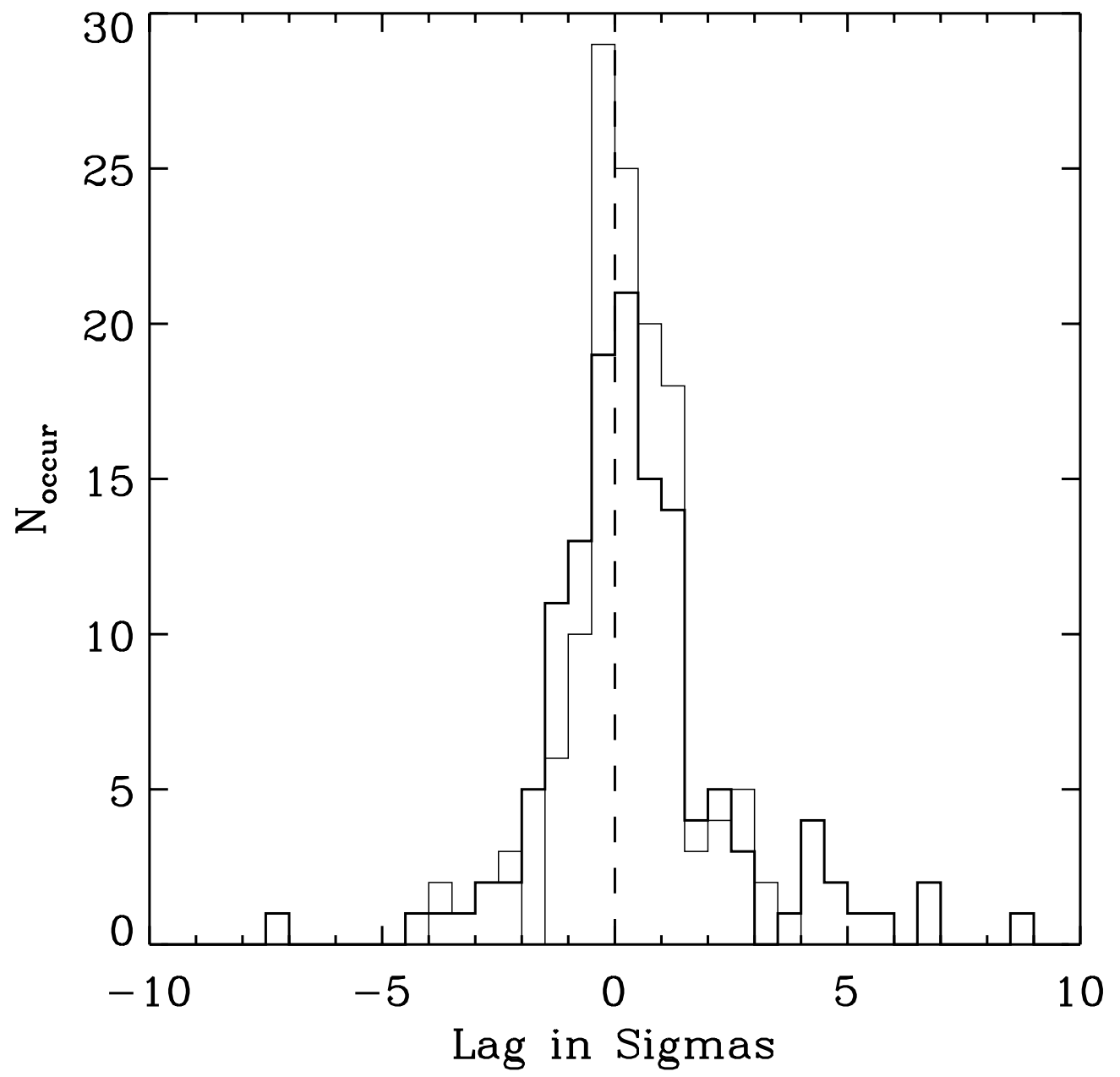

Fig. 4–Histograms of the lags for the original 260 short bursts of Figure 3 rendered in sigmas. Thick (thin) histogram is that half of the sample with $F_p$ > (<) 4.25 photons $cm^{-2}$ $s^{-1}$. Inspection of the time profiles reveals that the few negative outliers (< –2.5σ) appear to be valid, as do the larger number of positive outliers, which may represent the tail of the long burst population. A large majority (90–95%) are consistent with zero lag.

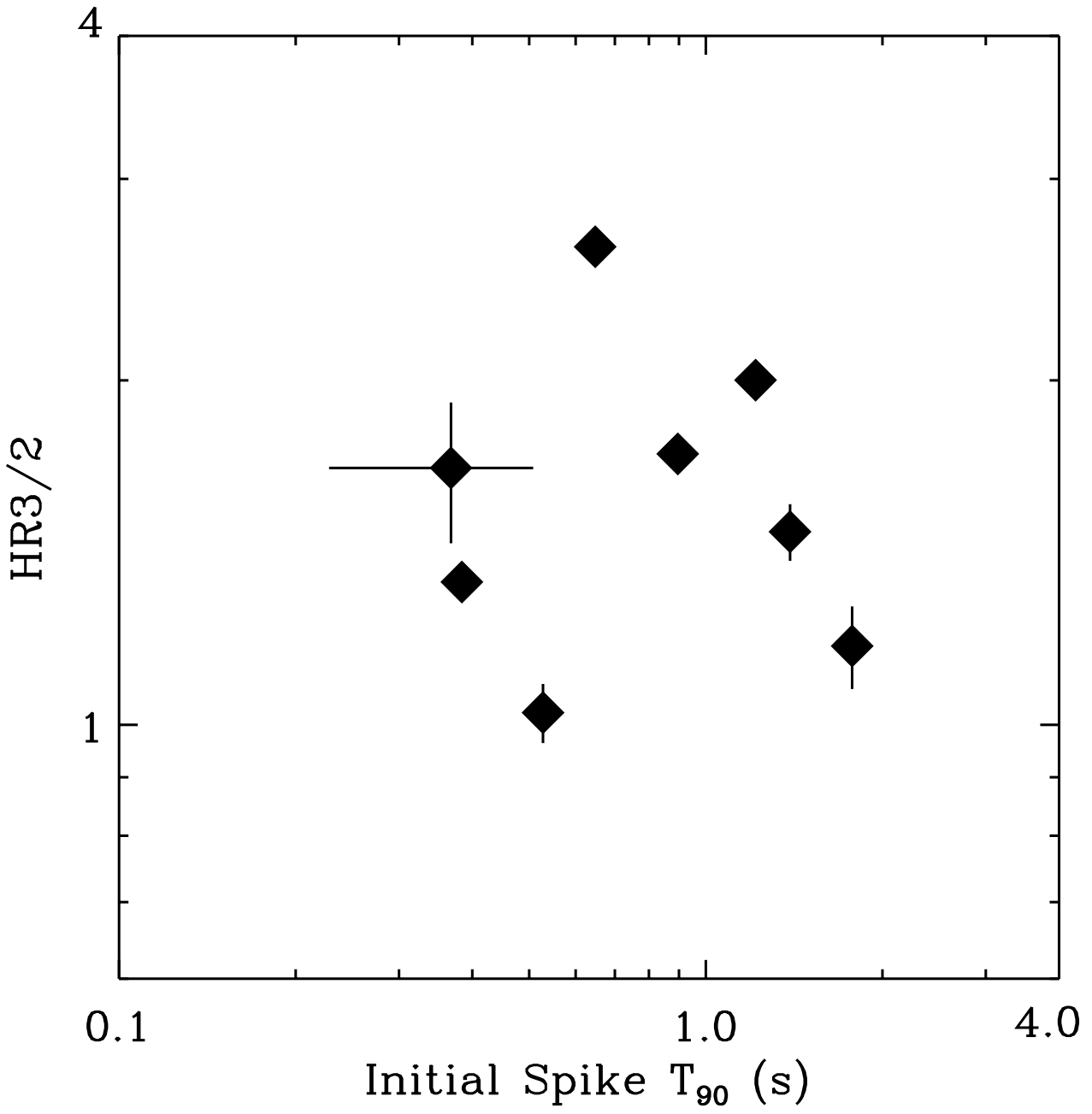

Fig. 5–The counts spectral hardness ratio, HR3/2 ([100–300 keV] / [25–50 keV]), of the spike emission for the eight bursts of Figure 2, vs. $T_{90}$ duration. Error bars (±1σ) are shown but in some cases are smaller than the plot symbols. The sample falls within the dynamic ranges for short bursts as reported in Kouveliotou et al. (1993).

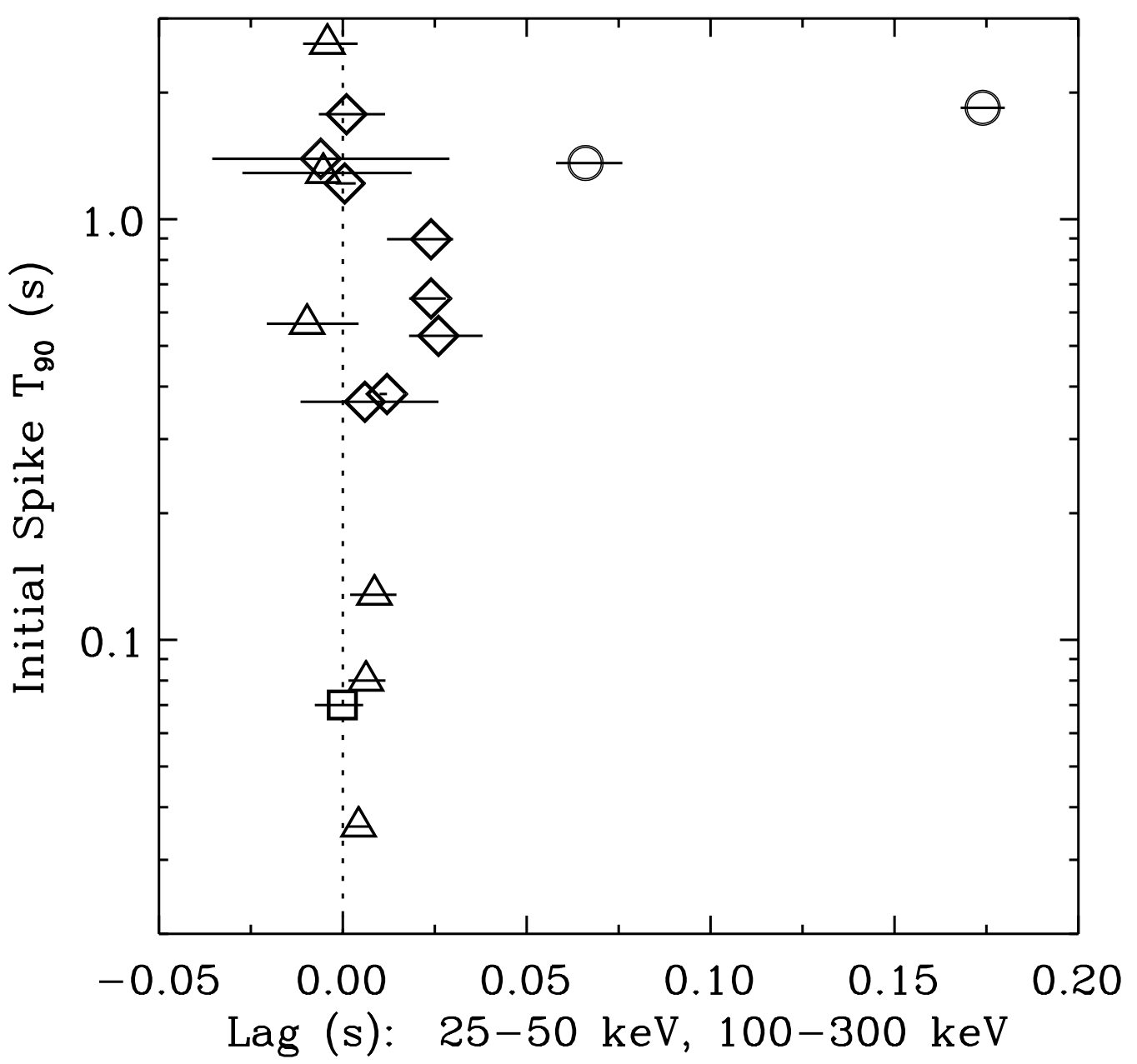

Fig. 6–$T_{90}$ duration vs. spectral lag for the spike emission in the eight BATSE bursts of Figure 2 (open diamonds), and the initial pulses of the two BATSE bursts in Figure 1 (open circles); both sets shown with ±1σ errors. The results for six Swift/BAT bursts are plotted (open triangles) with ±1σ errors. Also plotted is a measurement of the lag for the spike emission of the HETE-2 burst (open square, error bars: 3σ; see text), GRB 050709.

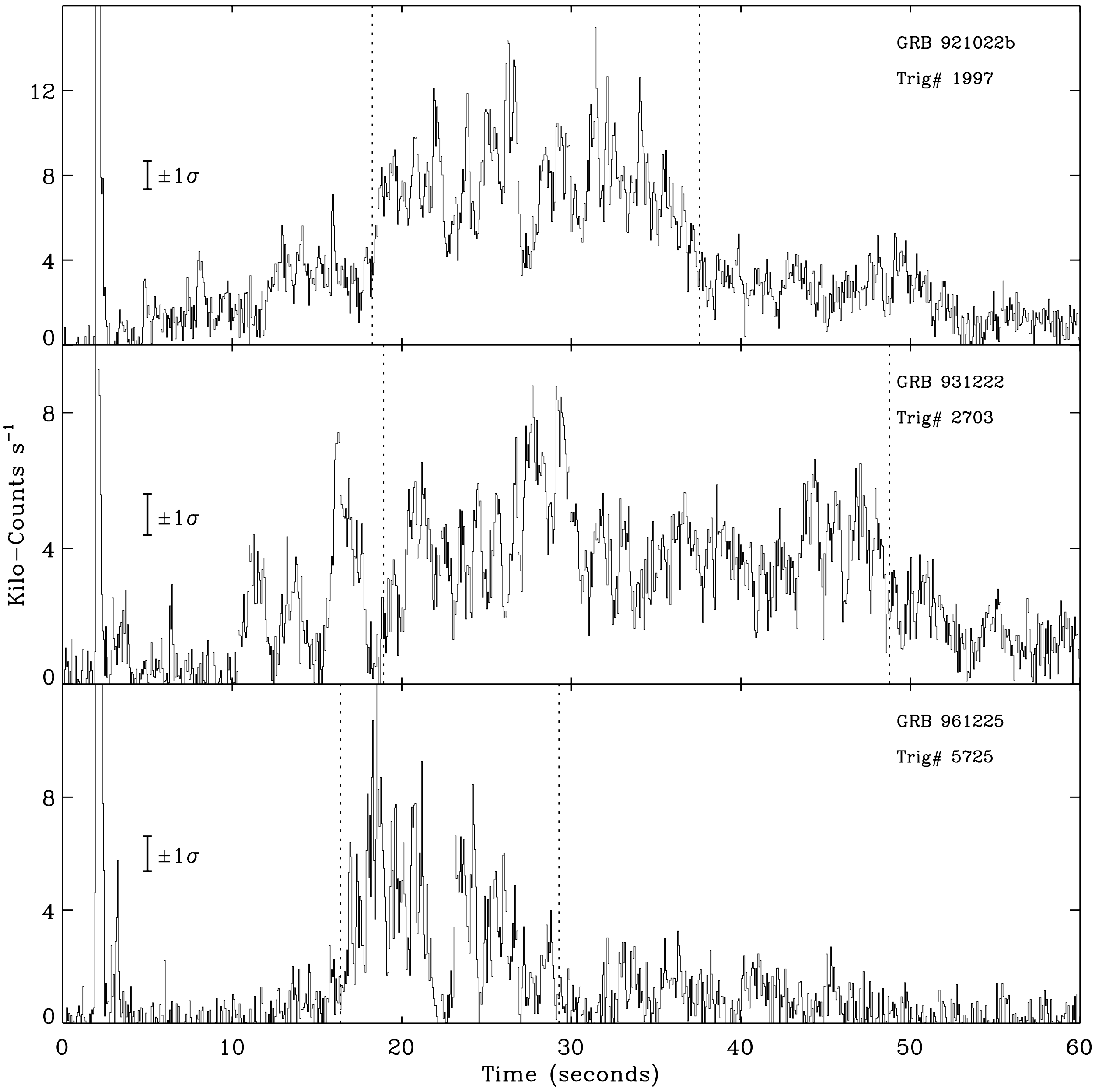

Fig. 7–Time profiles for the three bursts of Figure 2 with the brightest and most fluent extended emission. Vertical axis truncates initial spike to emphasize extended emission. Dotted lines indicate intervals (same as in Figure 2) analyzed for spectral evolution in § 2.2. Error bars to right of spikes represent $\pm 1\sigma$ fluctuations.

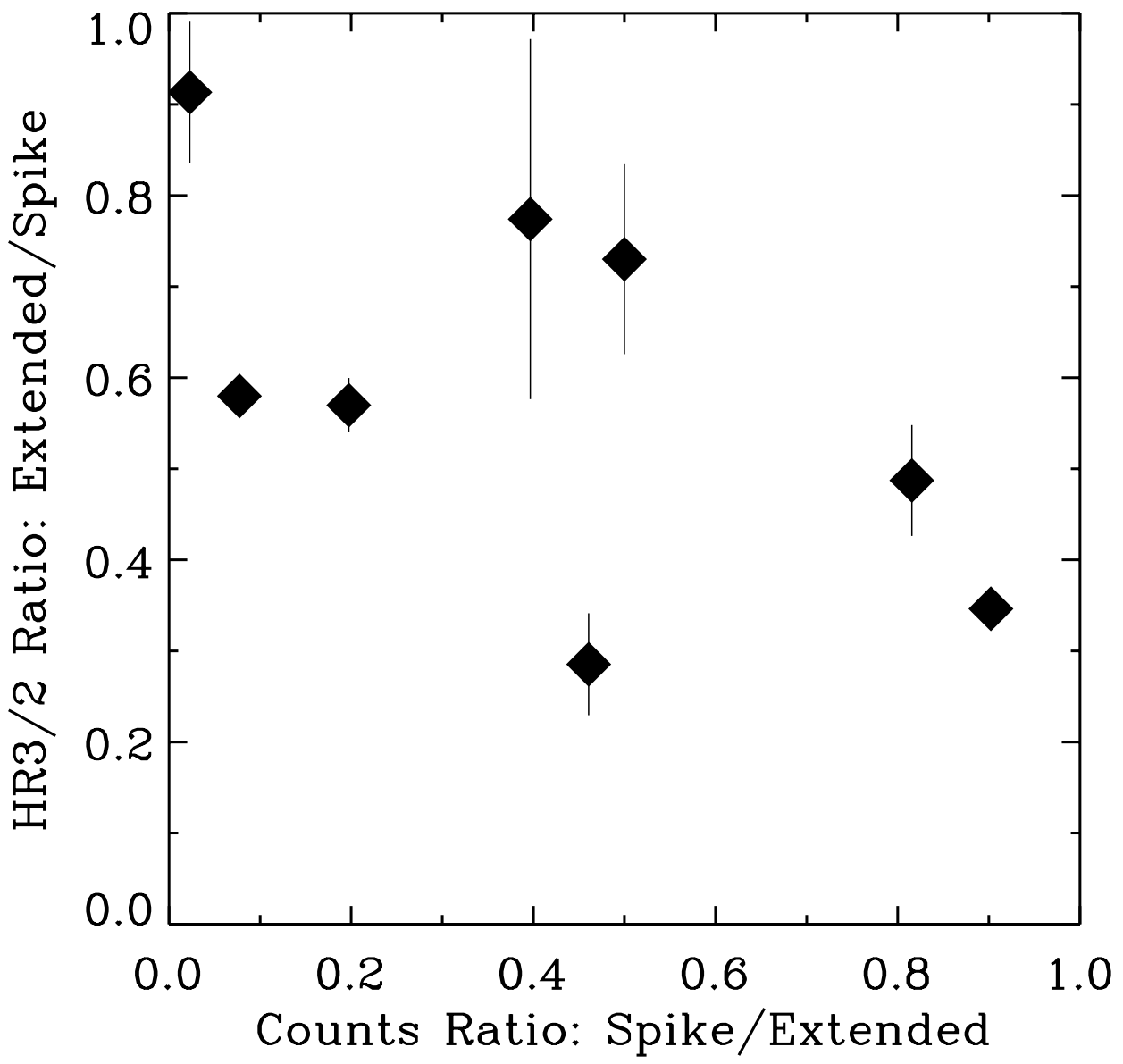

Fig. 8–Ratio of HR3/2 (spike : extended) vs. total counts (> 25 keV) ratio, extended : spike. There is a slight hint that, when the extended emission is relatively less fluent, it tends to be spectrally softer.